\documentclass[showpacs,preprintnumbers,amsmath,amssymb]{revtex4}

\usepackage{epsfig}
\usepackage{graphicx}
\usepackage{dcolumn}
\usepackage{bm}
\newcommand{\be}{\begin{equation}}

\newcommand{\ee}{\end{equation}}
\newcommand{\bea}{\begin{eqnarray}}
\newcommand{\eea}{\end{eqnarray}}

\begin{document}

\preprint{}

\title{Kernel Granger causality and the analysis of dynamical networks}
\date{\today}
\author{D.~Marinazzo}\email{daniele.marinazzo@ba.infn.it}
\author{M.~Pellicoro}\email{mario.pellicoro@ba.infn.it}
\author{S.~Stramaglia}\email{sebastiano.stramaglia@ba.infn.it}
\affiliation{Dipartimento  Interateneo di Fisica, Universit\`a di
Bari, I-70126 Bari, Italy,} \affiliation{TIRES-Center of Innovative
Technologies for Signal Detection and Processing, Universit\`a di
Bari,  Italy,} \affiliation{I.N.F.N., Sezione di Bari, I-70126 Bari,
Italy}

\date{\today}

\begin{abstract}
We propose a method of analysis of dynamical networks based on a
recent measure of Granger causality between time series, based on
kernel methods. The generalization of kernel Granger causality to
the multivariate case, here presented, shares the following features
with the bivariate measures: (i) the nonlinearity of the regression
model can be controlled by choosing the kernel function and (ii) the
problem of false-causalities, arising as the complexity of the model
increases, is addressed by a selection strategy of the eigenvectors
of a reduced Gram matrix whose range represents the additional
features due to the second time series. Moreover, there is no {\it a
priori} assumption that the network must be a directed acyclic
graph. We apply the proposed approach to a network of chaotic maps
and to a simulated genetic regulatory network: it is shown that the
underlying topology of the network can be reconstructed from time
series of node's dynamics, provided that a sufficient number of
samples is available. Considering a linear dynamical network, built
by  preferential attachment scheme, we show that for limited data
use of bivariate Granger causality is a better choice w.r.t methods
using $L1$ minimization. Finally we consider real expression data
from HeLa cells, 94 genes and 48 time points. The analysis of static
correlations between genes reveals two modules corresponding to well
known transcription factors; Granger analysis puts in evidence
nineteen causal relationships, all involving genes related to tumor
development. 
\end{abstract}

\maketitle

\section{Introduction}
Dynamical networks \cite{barabasi} model physical and biological
behavior in many applications; examples range from networks of
neurons \cite{netnn}, Josephson junctions arrays \cite{jose} to
genetic networks \cite{gennet}, protein interaction nets
\cite{protein} and metabolic networks \cite{metabolic}.
Synchronization in dynamical networks is influenced by the topology
of the network \cite{bocca}. A great need exists for the development
of effective methods of inferring network structure from time series
data; a method for detecting the topology of dynamical networks,
based on chaotic synchronization, has been proposed in \cite{yu1}; a
recent approach deals with the case of low number of samples and
proposed methods rooted on L1 minimization \cite{sauer}.

Granger causality has become the method of choice to determine
whether and how two time series exert causal influences on each
other \cite{hla}. This approach is based on prediction: if the
prediction error of the first time series is reduced by including
measurements from the second one in the linear regression model,
then the second time series is said to have a causal influence on
the first one. This frame has been used in many fields of science,
including neural systems \cite{ding}, reo-chaos \cite{reo} and
cardiovascular variability \cite{faes}. The estimation of linear
Granger causality from Fourier and wavelet transforms of time series
data has been recently addressed \cite{ding-prl}.

Kernel algorithms work by embedding data into a Hilbert space, and
searching for linear relations in that space \cite{vapnik}. The
embedding is performed implicitly, by specifying the inner product
between pairs of points \cite{shawe}. We have recently exploited the
properties of kernels to provide nonlinear measures of bivariate
Granger causality \cite{noi}. We reformulated linear Granger
causality and introduced a new statistical procedure to handle
over-fitting \cite{palus} in the linear case. Our new formulation
was then generalized to the nonlinear case by means of the kernel
trick \cite{shawe}, thus obtaining a method with the following two
main features: (i) the nonlinearity of the regression model can be
controlled by choosing the kernel function; (ii) the problem of
false-causalities, which arises as the complexity of the model
increases, is addressed by a selection strategy of the eigenvectors
of a reduced Gram matrix whose range represents the additional
features due to the second time series.

In this paper we describe in detail the kernel Granger approach and
address use of Granger causality to estimate, from multivariate time
series data, the topology and the drive-response relationships of a
dynamical network. To this aim, we generalize our method in
\cite{noi} to the case of multivariate data.

The paper is organized as follows. In the next section we describe
the kernel Granger causality method in the bivariate case, adding
some details to the presentation in \cite{noi}. In section III we
generalize the method to process multivariate time series data, and
show that the proposed method can discern whether an interaction is
direct or mediated by a third time series. The case of dynamical
networks is described in section IV: we analyze a system of
interacting chaotic maps and a model of gene regulatory networks. In
Section V we consider systems with sparse connectivity and limited
data, and compare the bivariate Granger approach with a multivariate
method based on L1 minimization; then we analyze a real data set,
gene expressions of HeLa cells. Section V summarizes our
conclusions.

\section{Bivariate Granger causality}
In this section we review the kernel method for Granger causality
proposed in \cite{noi}. Let us start with the linear case.
\subsection{Linear Granger causality}
Suppose to model the temporal dynamics of a stationary time series
$\{\xi_n\}_{n=1,.,N+m}$ by an autoregressive model of order $m$:
$$\xi_n=\sum_{j=1}^m A_j\; \xi_{n-j}+E_n,$$
and by a bivariate autoregressive model which takes into account
also a simultaneously recorded time series $\{\eta_n\}_{n=1,.,N+m}$:
$$\xi_n=\sum_{j=1}^m A^\prime_j \;\xi_{n-j}+\sum_{j=1}^m B_j \;\eta_{n-j}+E^\prime_n.$$
The coefficients of the models are calculated by standard least
squares optimization; $m$ is usually chosen according to Akaike
criterion \cite{akaike} or embedding techniques from the theory of
nonlinear dynamical systems \cite{kantz}.

The concept of Granger causality is  \cite{hla}: $\eta$
Granger-causes $\xi$ if the variance of residuals $E^\prime$ is
significantly smaller than the variance of residuals $E$, as it
happens when coefficients $B_j$ are jointly significantly different
from zero. This can be tested by performing an F-test or Levene's
test for equality of variances \cite{geweke}. An index measuring the
strength of the causal interaction is then defined as:
\begin{equation}\label{causalityindex}
\delta=1-{\langle {E^\prime}^2\rangle\over \langle E^2\rangle},
\end{equation}
where $\langle \cdot\rangle$ means averaging over $n$ (note that
$\langle E\rangle =  \langle E^\prime \rangle =0$). Exchanging the
roles of the two time series, one may equally test causality in the
opposite direction, i.e. to check whether $\xi$ Granger-causes
$\eta$.

We  use the shorthand notations:
$$X_i=(\xi_{i},\ldots,\xi_{i+m-1})^\top,$$
$$Y_i=(\eta_{i},\ldots,\eta_{i+m-1})^\top$$ and $x_i=\xi_{i+m}$, for
$i=1,\ldots,N$. We treat these quantities as $N$ realizations of the
stochastic variables $X$, $Y$ and $x$. Let us denote  $\bf{X}$ the
$m\times N$ matrix having vectors $X_i$ as columns, and $\bf{Z}$ the
$2m\times N$ matrix having vectors $Z_i = (X_i^\top, Y_i^\top)^\top$
as columns.  The values of $x$ are organized in a vector
$\bf{x}$=$(x_1,\ldots,x_N)^\top$. In full generality we assume that
each component of $X$ and $Y$ has zero mean, and that vector
$\bf{x}$ has zero mean and is normalized, i.e. $\bf{x}^\top
\bf{x}$=$1$.

Now, for each $i=1,\ldots,N$, we define:
$$\tilde{x_i}=\sum_{j=1}^m A_j\; \xi_{i+m-j},$$
$$\tilde{x_i}^\prime=\sum_{j=1}^m A^\prime_j \;\xi_{i+m-j}+\sum_{j=1}^m B_j \;\eta_{i+m-j}.$$
The vectors $\tilde{\bf{x}}$=$(\tilde{x_1},\ldots,\tilde{x_N})^\top$
and
$\tilde{\bf{x}}^\prime$=$(\tilde{x_1}^\prime,\ldots,\tilde{x_N}^\prime)^\top$
are the estimated values by linear regression, in the two cases. It
is easy to show that $\tilde{\bf{x}}$ and $\tilde{\bf{x}}^\prime$
have the following geometrical interpretation. Let $H\subseteq
\Re^N$ be the range of  the $N\times N$ matrix
$\textbf{K}=\textbf{X}^\top \textbf{X}$. Then $\tilde{\bf{x}}$ is
the projection of $\bf{x}$ on $H$. In other words, calling
$\bf{v}_1,\ldots,\bf{v}_{m}$ the (orthonormal) eigenvectors of
$\textbf{K}$ with non-vanishing eigenvalue and  $$P=\sum_{i=1}^m
\bf{v}_i \bf{v}_i^\top$$ the projector on the space $H$, we have
$\tilde{\bf{x}}=P \bf{x}$. Let us define $\bf{y}=\bf{x}-P\bf{x}$.
Analogously $\tilde{\bf{x}}^\prime=P^\prime\bf{x}$, $P^\prime$ being
the projector on the $2m$-dimensional space $H^\prime\subseteq
\Re^N$, equal to the range of the matrix
$\textbf{K}^\prime=\textbf{Z}^\top \textbf{Z}$. Moreover, it is easy
to show that
\begin{equation}\label{causalityindex1}
\delta={    {\tilde{\bf{x}}}^{\prime^\top}\tilde{\bf{x}}^\prime
-\tilde{\bf{x}}^\top \tilde{\bf{x}}    \over 1- \tilde{\bf{x}}^\top
\tilde{\bf{x}} }.
\end{equation}
Now note that $H\subseteq H^\prime$. Therefore we may decompose
$H^\prime$ as follows: $H^\prime=H\oplus H^\perp$, where $H^\perp$
is the space of all vectors of $H^\prime$ orthogonal to all vectors
of $H$.  $H^\perp$ corresponds to the additional features due to the
inclusion of $\{\eta\}$ variables. Calling $P^\perp$ the projector
on $H^\perp$, we can write:
\begin{equation}\label{causalityindex2}
\delta={||P^\perp \bf{y}||^2\over 1-\tilde{\bf{x}}^\top
\tilde{\bf{x}}}.
\end{equation}
Now we note that $H^\perp$ is the range of the matrix
$$\tilde{\bf{K}}=\bf{K}^\prime-\bf{K}^\prime P-P\left(\bf{K}^\prime-\bf{K}^\prime P\right)=\bf{K}^\prime-P\bf{K}^\prime-\bf{K}^{\prime}P+P\bf{K}^{\prime}P.$$
Indeed, for any ${\bf u}\in \Re^N$, we have $\tilde{\bf{K}}{\bf
u}={\bf v}-P{\bf v}$, where ${\bf v}= \bf{K}^\prime\left({\bf I}-
P\right){\bf u}\in H^\prime$, and $\tilde{\bf{K}}{\bf u} \in
H^\perp$. It follows that $H^\perp$ is spanned by the set of the
eigenvectors, with non vanishing eigenvalue, of $\tilde{\bf{K}}$.
Calling $\bf{t}_1,\ldots,\bf{t}_{m}$ these eigenvectors, we have:
\begin{equation}\label{causalityindex3}
\delta=\sum_{i=1}^m r_i^2,
\end{equation}
where $r_i$ is the Pearson's correlation coefficient of $\bf{y}$ and
$\bf{t}_i$ (since the overall sign of $\bf{t}_i$ is arbitrary, we
can assume that $r_i$ is positive). Let $\pi_i$ be the probability
that $r_i$ is due to chance, obtained by Student's t test. Since we
are dealing with multiple comparison, we use the Bonferroni
correction to select the eigenvectors $\bf{t}_{i^\prime}$,
correlated with $\bf{y}$, with expected fraction of false positive
 q (equal to $0.05$). Therefore we calculate a new causality
index by summing, in equation (4), only over the $\{r_{i^\prime}\}$
such that $\pi_{i^\prime}<q/m$, thus obtaining a {\it filtered}
linear Granger causality index:
\begin{equation}\label{causalityindex4}
\delta_F={\sum_{i^\prime} r_{i^\prime}^2}\;\;.
\end{equation}
It is assumed that $\delta_F$ measures the causality $\eta \to \xi$,
without further statistical test.

\subsection{Kernel Granger causality} In this subsection  we describe
the generalization of linear Granger causality to the nonlinear
case, using methods from the theory of Reproducing Kernel Hilbert
Spaces (RKHS) \cite{shawe}. Given a kernel function $K$, with
spectral representation $K(X,X^\prime)=\sum_a \lambda_a \phi_a
(X)\phi_a (X^\prime)$ (see Mercer's theorem \cite{vapnik}), we
consider $H$, the range of the $N\times N$ Gram matrix $\bf{K}$ with
elements $K(X_i,X_j)$. In order to make the mean of all variables
$\phi_a(X)$ zero, we replace ${\bf K}\to {\bf K}-P_0{\bf K}-{\bf
K}P_0+P_0{\bf K}P_0$, where $P_0$ is the projector onto the
one-dimensional subspace spanned by the vector such that each
component is equal to the unity \cite{shawe}; in the following we
assume that this operation has been performed on each Gram matrix.
As in the linear case, we calculate $\tilde{\bf{x}}$, the projection
of $\bf{x}$ onto $H$. Due to the spectral representation of $K$,
$\tilde{\bf{x}}$ coincides with the linear regression of $\bf{x}$ in
the feature space spanned by $\sqrt{\lambda_a}\phi_a$, the
eigenfunctions of $K$; the regression is nonlinear in the original
variables.

While using both $X$ and $Y$ to predict $x$, we evaluate the Gram
matrix $\bf{K^\prime}$ with elements $K^\prime_{ij}=K(Z_i,Z_j)$. The
regression values now form the vector $\tilde{\bf{x}}^\prime$ equal
to the projection of $\bf{x}$ on $H^\prime$, the range of
$\bf{K^\prime}$.  Before evaluating the filtered causality index, as
in the linear case, we note that not all kernels may be used to
evaluate Granger causality. Indeed if $Y$ is statistically
independent of $X$ and $x$, then $\tilde{\bf{x}}^\prime$ and
$\tilde{\bf{x}}$ should coincide in the limit $N\to\infty$. This
property, invariance of the  risk minimizer  when statistically
independent variables are added to the set of input variables, is
satisfied only by suitable kernels, as discussed in \cite{ancona}.
In the following we consider two possible choices, which fulfill the
invariance requirement.

\noindent {\it Inhomogeneous polynomial kernel}. The  inhomogeneous
polynomial (IP) kernel of integer order $p$ is:
$$ K_p(X,X^\prime)=\left(1+X^\top X^\prime\right)^p.$$ In this case
the eigenfunctions are made of all the monomials, in the input
variables, up to the $p-th$ degree. The dimension of the space $H$
is $m_1=1/B(p+1,m+1) -1$, where $B$ is the beta function, and $p=1$
corresponds to the linear regression. The dimension of space
$H^\prime$ is $m_2=1/B(p+1,2m+1) -1$. As in the linear case, we note
that $H\subseteq H^\prime$ and decompose $H^\prime=H\oplus H^\perp$.
Subsequently we calculate
$\tilde{\bf{K}}=\bf{K}^\prime-P\bf{K}^\prime-\bf{K}^{\prime}P+P\bf{K}^{\prime}P$;
the dimension of the range of $\tilde{\bf{K}}$ is $m_3=m_2 -m_1$.
Along the same lines as those described in the linear case, we
construct the kernel Granger causality taking into account only the
eigenvectors of $\tilde{\bf{K}}$ which pass the Bonferroni test:
\begin{equation}\label{causalityindex5}
\delta_F^K={\sum_{i^\prime} r_{i^\prime}^2},
\end{equation}
the sum being only over the eigenvectors of $\tilde{\bf{K}}$ with
probability $\pi_{i^\prime}<q/m_3$.

\noindent {\it Gaussian kernel}. The Gaussian kernel reads:
\begin{equation}
K_\sigma(X,X^\prime)=\exp{\left(-{\left(X-X^\prime\right)^\top
\left(X-X^\prime\right)\over 2\sigma^2}\right)}, \label{gau}
\end{equation}
and depends on the width $\sigma$. $\sigma$ controls the complexity
of the model: the dimension of the range of the Gram matrix
decreases as $\sigma$ increases. As in previous cases, we may
consider $H$, the range of the Gram matrix $\bf{K}$, and $H^\prime$,
the range of $\bf{K^\prime}$, but in this case the condition
$H\subseteq H^\prime$ would not necessarily hold; therefore some
differences in the approach must be undertaken. We call $L$ the
$m_1$-dimensional span of the eigenvectors of $\bf{K}$ whose
eigenvalue is not smaller than $\mu \lambda_{max}$, where
$\lambda_{max}$ is the largest eigenvalue of $\bf{K}$ and $\mu$ is a
small number (we use $10^{-6}$). We evaluate $\tilde{\bf{x}}=P{\bf
x}$, where $P$ is the projector on $L$. After evaluating the Gram
matrix $\bf{K}^\prime$, the following matrix is considered:
\begin{equation} {\bf K}^*=\sum_{i=1}^{m_2} \rho_i {\bf w}_i {\bf
w}_i^\top, \label{kstar}\end{equation} where $\{{\bf w}\}$ are the
eigenvectors of $\bf{K}^\prime$, and the sum is over the eigenvalues
$\{ \rho_i\}$ not smaller than $\mu$ times the largest eigenvalue of
$\bf{K}^\prime$. Then we evaluate
$\tilde{\bf{K}}=\bf{K}^*-P\bf{K}^*-\bf{K}^*P+P\bf{K}^*P$, and denote
$P^\perp$  the projector onto the $m_3$-dimensional range of
$\tilde{\bf{K}}$. Note that the condition $m_2=m_1+m_3$ may not be
strictly satisfied in this case (however in our experiments we find
that the violation of this relation is always very small, if any).
The kernel Granger causality index for the Gaussian kernel is then
constructed as in the previous case, see equation
(\ref{causalityindex5}).

\section{Multivariate kernel causality}
Let $\{\xi(a)_n\}_{n=1,.,N+m}$, $a=1,\ldots,M$, be $M$
simultaneously recorded time series. In order to put in evidence the
drive-response pattern in this system, one may evaluate the
bivariate Granger causality, described in the previous sections,
between every pair of time series. It is recommended, however, to
treat the data-set as a whole, thus generalizing kernel Granger
causality to the multivariate case, as follows. We denote
$$X(c)_i=(\xi(c)_{i},\ldots,\xi(c)_{i+m-1})^\top ,$$ for $c=1,\ldots,M$
and $i=1,\ldots,N$. In order to evaluate the causality
$\{\xi(a)\}\to \{\xi(b)\}$, we define, for $i=1,\ldots,N$, $$Z_i =
(X(1)_i^\top,\ldots,X(a)_i^\top, \ldots,X(M)_i^\top )^\top ,$$
containing all the input variables, and $$X_i =
(X(1)_i^\top,\ldots,X(M)_i^\top )^\top ,$$ containing all the input
variables but those related to $\{\xi(a)\}$. Gram matrices $\bf{K}$
and $\bf{K}^\prime$ are then evaluated: $K_{ij}=K(X_i,X_j)$ and
$K^\prime_{ij}=K(Z_i,Z_j)$. The target vector is now
$\bf{x}$=$(\xi(b)_{1+m},\ldots,\xi(b)_{N+m})^\top$. Along the same
lines as in the bivariate case, for IP kernel or the Gaussian one,
we then calculate the causality index as in equation (6): it is
denoted $\delta_F^K\left(a\to b\right)$ and measures the strength of
causality $a\to b$, taking into account all the available variables.
Repeating these steps for all $a$ and $b$, the casuality pattern in
the data set is evaluated. Note that the threshold for the
Bonferroni's correction, in the multivariate case, must be lowered
by the number of pairs $M(M-1)/2$.
\subsection{Three coupled maps}
\begin{figure}[ht]
\includegraphics[width=10cm]{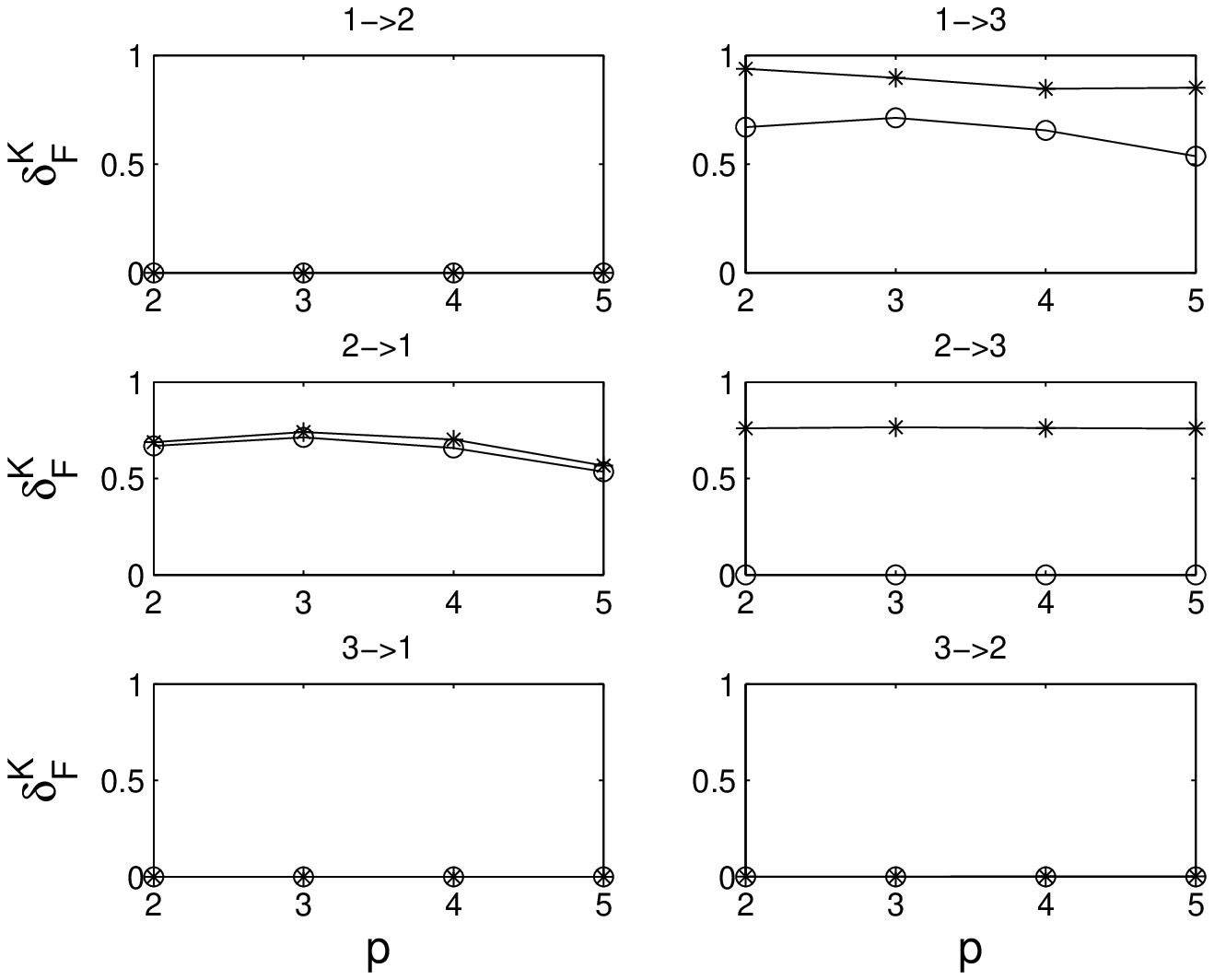}\caption{{\rm  The causal
relationships between all pairs of maps, in the  example of three
logistic maps described in the text, using multivariate Granger
causality (empty circles) and bivariate Granger causality (stars).
Here $m=1$ and the IP kernel, with various values of $p$, is used.
\label{fig1}}}\end{figure}

As an example, we consider the following system of three logistic
maps \cite{ott}:
\begin{eqnarray}
\begin{array}{l}
x_{1,t}=0.8\left(1-a x_{1,t-1}^2\right)+0.2\left(1-a x_{2,t-1}^2\right)+s \tau_{1,t},\\
x_{2,t}=1-a x_{2,t-1}^2 +s \tau_{2,t},\\
x_{3,t}=0.8\left(1-a x_{3,t-1}^2\right)+0.2\left(1-a
x_{1,t-1}^2\right)+s \tau_{3,t};
\end{array}
\label{map}
\end{eqnarray}
here $a=1.8$, $s=0.01$ and $\tau$'s are unit variance Gaussian noise
terms. The causal relationships implemented in these equations are
$2 \to 1$ and $1\to 3$. Analyzing segments of length $N=1000$, we
evaluate both the multivariate causality, as described in section 4,
and the bivariate causality for all pairs of maps. We use the IP
kernel with various values of $p$; the results are displayed in
figure (\ref{fig1}). It is well known \cite{ding} that performing
pairwise evaluation for multivariate data has the drawback that one
cannot discern whether the influence between two time series is
direct or is mediated by other time series. This is what happens in
the present example. Both the multivariate and the bivariate
analysis reveal the influences $2 \to 1$ and $1\to 3$. On the other
hand, the bivariate analysis reveals also the influence $2\to 3$,
which is actually mediated by $1$: the multivariate analysis
recognizes $2\to 3$ as non-significative.

\section{Analysis of dynamical networks}
In this section we simulate two dynamical networks and apply the
multivariate Granger analysis to estimate the topology structure of
 systems from time series data.
\subsection{Network of chaotic maps}
\begin{figure}[ht]
\includegraphics[width=10cm]{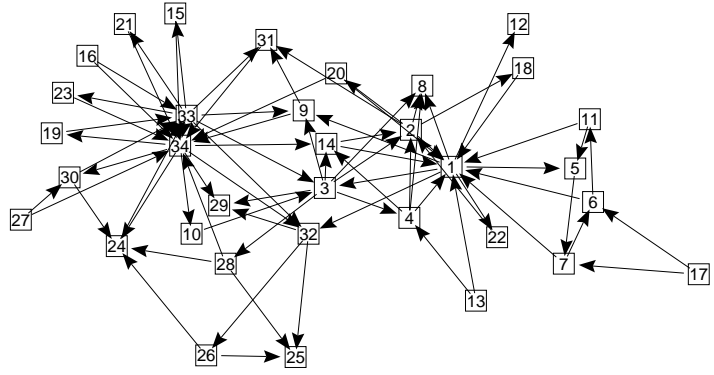}\caption{{\rm  The
directed network of 34 nodes obtained assigning randomly a direction
to links of the Zachary network. \label{fig2}}}\end{figure}

Let us consider a coupled map lattice of $n$ nodes, with equations,
for $i=1,\ldots,n$:
\begin{equation}
x_{i,t}=\left(1-\sum_{j=1}^n c_{ij}\right)\left(1-a
x_{i,t-1}^2\right)+\sum_{j=1}^n c_{ij}\left(1-a x_{j,t-1}^2\right)+s
\tau_{i,t}, \label{cml}
\end{equation}
where $a$, $s$ and $\tau$'s are the same as in equation (\ref{map}),
and $c_{ij}$ represents the coupling $j\to i$. We fix $n=34$ and
construct couplings as follows. We consider the  well known Zachary
data set  \cite{zachari}, an undirected network of $34$ nodes. We
assign a direction to each link, with equal probability, and set
$c_{ij}$ equal to 0.05, for each link of the directed graph thus
obtained, and zero otherwise. The network is displayed in figure
(\ref{fig2}): here the goal is to estimate this directed network
from the measurements of time series on  nodes.

\begin{figure}[ht]
\includegraphics[width=10cm]{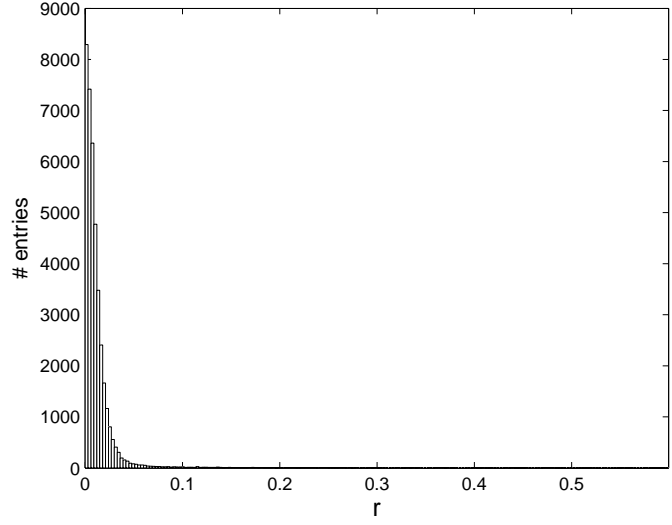}\caption{{\rm The
distribution of the 39270 $r$-values calculated while evaluating the
causality indexes of the coupled map lattice (see the text).
\label{fig3}}}\end{figure}

\begin{figure}[ht]
\includegraphics[width=10cm]{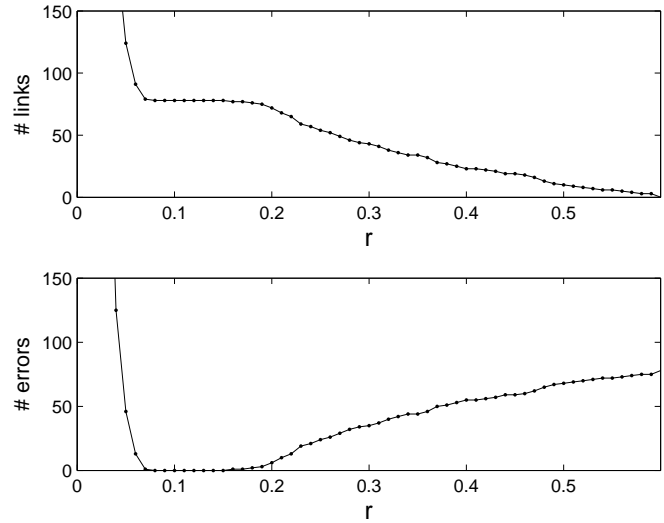}\caption{{\rm (top) Concerning the coupled map lattice, the
horizontal axis represents the threshold for the values of $r$; the
plot shows the number of links of the directed network constructed
from the projections whose Pearson's coefficient exceeds the
threshold. (bottom) The total number of errors, in the reconstructed
network, is plotted versus the threshold $r$. At large $r$ the
errors are due only to missing links, whereas at large $r$ the
errors are due only to links that do not exist in the true network
and are recovered. \label{fig4}}}\end{figure}

\begin{figure}[ht]
\includegraphics[width=10cm]{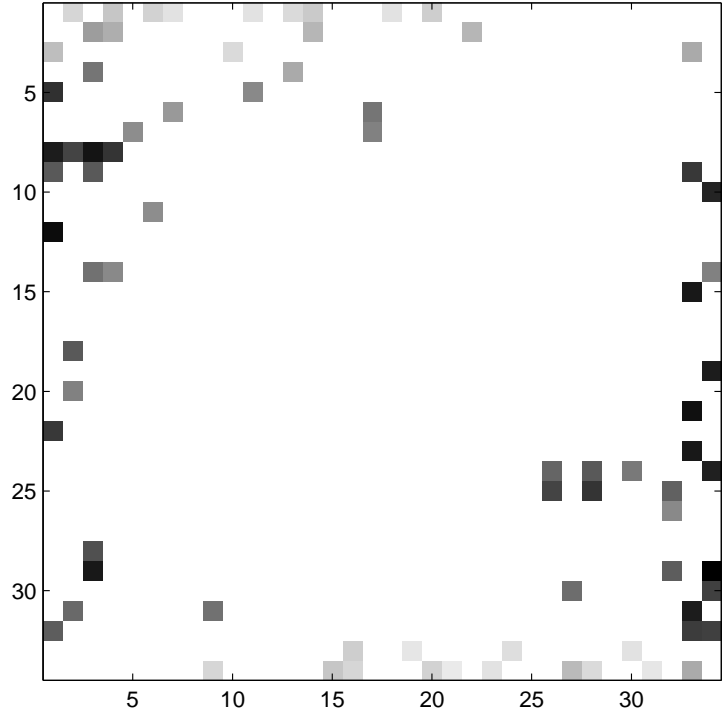}\caption{{\rm The causality
indexes $\delta_F^K (j\to i)$, for all pairs of maps, are
represented in a gray scale. White square means zero causality,
black squares correspond to the maximal causality observed (0.55).
\label{fig5}}}\end{figure}

The multivariate Granger analysis, described in the previous
section, perfectly recovers the underlying network using the  IP
kernel with $p=2$, $m=1$ and $N=10000$. Note that while evaluating
$\delta_F^K (j\to i)$, for all $i$ and $j$, 39270 Pearson's
coefficients $r$ are calculated. Their distribution is represented
in figure (\ref{fig3}): there is a strong peak at $r=0$
(corresponding to projections that are discarded), and a very low
number of projections with a rather large value of $r$, up to about
$r=0.6$. It is interesting to describe the results in terms of a
threshold for correlations. Given a  threshold value $r$, we select
the correlation coefficients whose value is greater than $r$. We
then calculate the corresponding causality indexes $\delta_F^K (j\to
i)$, and construct the directed network whose links correspond to
non-vanishing elements of $\delta_F^K (j\to i)$. In figure
(\ref{fig4})-top we plot the total number of links of the
reconstructed network, as a function of the threshold $r$: the curve
shows a plateau, around $r=0.1$, corresponding to a directed network
which is stable against variations of  $r$. At the plateau $428$
projections are selected, which coincide with those selected by
means of Bonferroni's test. In figure (\ref{fig4})-bottom we plot
the number of errors (the sum of the number of links that exist in
the true network and are not recovered, plus the number of links
that exist only in the recovered network) versus the threshold $r$:
the plateau leads to perfect reconstruction of the original network.
We stress that a large number of samples is needed to recover the
underlying network: in the typical case we find that the network is
perfectly reconstructed if $N \ge 5000$, whilst if $N$ is further
lowered some errors in the reconstructed network appear. Moreover,
it is important to observe that, although all couplings $c_{ij}$
have the same magnitude, the causality strengths $\delta_F^K (j\to
i)$ depend on the link, as it is shown in figure (\ref{fig5}).
Granger causality is a measure of the information being transferred
from one time series to another, and it should not be identified
with couplings.

\subsection{Genetic regulatory network}
In this subsection we consider time series from a model of genetic
regulatory network  made of genes linked by weighted connections
(inhibitory or excitatory) \cite{yu}. The expression levels of all
genes, organized in a vector ${\bf g}$, evolve as follows:
\begin{equation}\label{genes}
{\bf g}_{t+1}={\bf g}_{t}+A\left({\bf g}_{t}-T {\bf I}\right) +{\bf
\Sigma},
\end{equation}
where $A$ is a connectivity strength matrix corresponding to the
network, $T=50$, ${\bf I}$ is the identity matrix and ${\bf \Sigma}$
is a vector of random variables uniform in $[-10,10]$. The values of
${\bf g}$ are restricted by floor and ceiling function to range in
$[0,100]$: this constraint provides the nonlinear character of the
model. As the simulation runs, multivariate data are sampled every
$t_s$ time steps. Moreover, the continuous data values are
discretized into $n_c$ categories (with equal bin sizes). In
\cite{yu} dynamic Bayesian network (DBN) models \cite{friedman} were
trained to data of length $N$ to recover the structure of matrix
$A$: the values $t_s=5$, $n_c =3$ and $N\ge 2000$ were found to lead
to the best reconstruction of genetic networks by DBN.

\begin{figure}[ht]
\includegraphics[width=10cm]{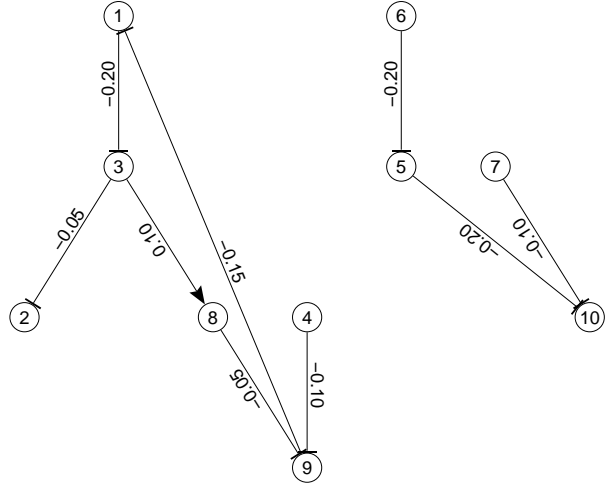}\caption{{\rm The genetic regulatory network analyzed in the text. Numbers next to
links specify regulation strength; arrows: excitatory; flat heads:
inhibitory.\label{fig6}}}\end{figure}

\begin{figure}[ht]
\includegraphics[width=10cm]{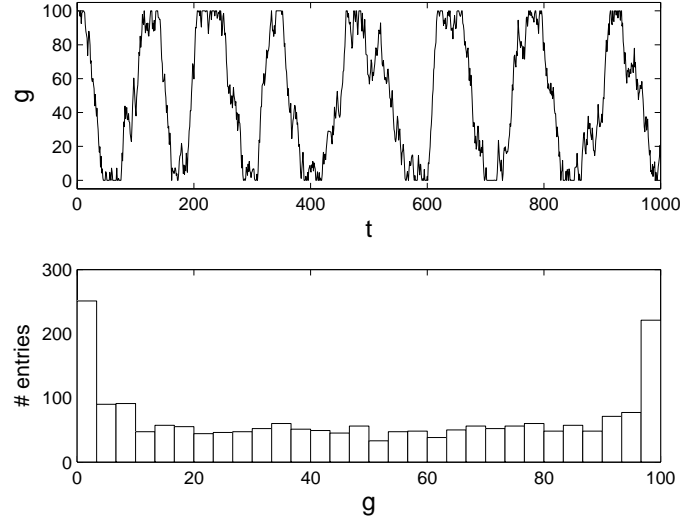}\caption{{\rm Top: the typical curve of expression  of a gene in the simulated regulatory network. Bottom: the distribution of expressions for the same gene. \label{fig7}}}\end{figure}

The genetic network we consider here is an example from \cite{yu}
and consists of ten genes with connections described in figure
(\ref{fig6}): there are two independent regulatory pathways, one of
which includes a large feedback structure. In figure (\ref{fig7})
the typical curve of expression for a gene in the network is
represented (top); the distribution of expressions, for the same
gene, is bimodal (bottom). We simulate 100 times the system
equations, starting from different initial conditions, and sample
time series of length $N=2000$, $t_s=5$ and $n_c=3$. The typical
recovered network by DBN, on this example, corresponds to one
missing link (from node 7 to 10) \cite{yu}. The linear multivariate
Granger approach, with $m=2$, leads to perfect reconstruction of the
network in 90 cases out of the 100; we obtain similar results on all
the examples presented in \cite{yu}. Using IP (with $p > 1$) and
Gaussian kernels we obtain similar performances as the ones from the
linear kernel.

\begin{figure}[ht]
\includegraphics[width=10cm]{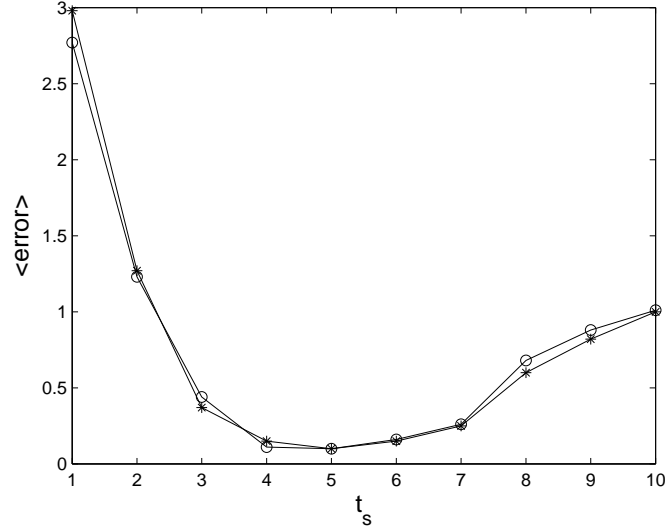}\caption{{\rm The mean
number of  errors (over 100 realization of the system of length
$N=2000$) obtained using linear kernel ($p=1$, empty circles) and
Gaussian kernel (stars), is plotted versus the sampling time $t_s$.
In the Gaussian case, $\sigma =50$ is used. The level of
discretization is $n_c=3$ and $m=2$. \label{fig8}}}\end{figure}

It is interesting to stress that the possibility that one has to
reconstruct the true genetic network depends on the sampling rate.
In figure (\ref{fig8}) we plot the mean number of errors (over 100
realizations) in the reconstructed network as a function of $t_s$
for the linear kernel and Gaussian kernel: the performance degrades
as $t_s$ gets far from 5.

Let us now discuss the case of large $t_s$:  all Granger causalities
are recognized as non significative. On the other hand, at large
$t_s$, we find significant {\it static} linear correlations between
time series of all pairs of genes belonging to the same pathway. In
other words, referring to figure (\ref{fig6}), the linear
correlations of times series from every pair of genes extracted from
$\{1,2,3,4,8,9\}$ is significant, as well as the linear correlation
for every pair extracted from $\{5,6,7,10\}$; consistently the
linear correlation, for pairs of genes from different pathways, is
not significant. We conclude that, at large $t_s$ Granger causality
analysis is not effective, but static correlation analysis may
anyway put in evidence groups of genes belonging to the same
regulatory pathway.

\section{Sparsely connected dynamical networks and limited data}
The Granger causality approach for dynamical networks here presented
requires a large amount of data samples to provide trustable
answers. However, there are situations (frequent in Bioinformatics)
where the number of samples is smaller than the number of variables
(genes): in these situations the multivariate Granger approach is
unfeasible. Under the assumption of sparse connectivity, it has been
proposed to replace least squares methodology with a multivariate
approach using  minimization with respect to $L1$ norm \cite{sauer}.
Here we show that there are situations where, even in presence of
sparse connectivity and limited data, use of bivariate Granger
causality is a better choice w.r.t $L1$ minimization. Indeed, in
these cases, the statistical robustness of the estimation of
information flow between pairs of time series may still be good,
with the drawback that some causality links, found by the bivariate
approach, may not be direct but mediated.

\begin{figure}[ht]
\includegraphics[width=10cm]{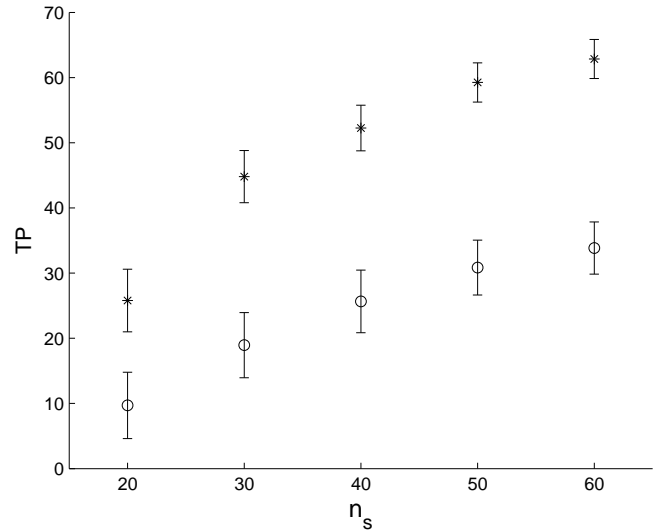}\caption{{\rm The true positive connections
found by bivariate Granger approach (stars) and by multivariate $L1$
minimization (empty circles)on the preferential attachment dynamical
network. The error bar represents one standard deviation, evaluated
over 50 realizations. The number of false positive connections is
always 5. \label{fig9}}}\end{figure}

We construct a network of $100$ nodes and $100$ links using the
preferential attachment procedure \cite{pa}; we give randomly a
direction to each link, with $1/2$ probabilities, thus obtaining a
directed network. Let us denote $d(i)$  the number of nodes from
which a link pointing to $i$ starts. We evolve a linear system on
this network, with equations:
\begin{equation}
x_{i,t}= a_i \;x_{i,t-1}+\sum_{j\to i}\left({0.8\over d(i)}\right)
x_{j,t-1}+ \tau_{i,t}. \label{lll}
\end{equation}
The sum is over nodes such that $j\to i$ is a link of the network;
$\tau$'s are unit variance Gaussianly distributed noise terms;
 $a_i$ is one, if $d(i)=0$, and $0.2$ otherwise. After a
transient, we sample $n_s$ consecutive time points, with
$n_s=20,30,40,50,60$. The $L1$ approach we use is the following. For
each $i=1,\ldots,100$, we find the vector $\mathbf{c}$ with minimum
$L1$ norm, among all those satisfying:
\begin{equation}
x_{i,t+1}=\sum_{j=1}^{100} c_j x_{j,t}, \label{llll}
\end{equation}
$t=1,\ldots,n_s-1$. The interaction $j\to i$ is considered
significant if the absolute value of $c_j$ exceed a threshold, fixed
so that the total number of false positive connections is five.
Subsequently we apply the bivariate linear Granger approach,
described in Section IIA, for each pair of nodes:  also for
Granger's approach we fix a threshold for the correlation
coefficients $r$, see equation (\ref{causalityindex3}), so that the
total number of false positive connections is five. In figure
(\ref{fig9}) we depict the number of true positive connections found
by the two approaches, as a function of $n_s$. It is clear that here
the bivariate Granger approach outperforms $L1$ minimization.

\subsection{HeLa gene expression regulatory network}
HeLa is the most famous cell culture line to date \cite{hela}. These
are cells isolated from a human uterine cervical carcinoma in 1951
and used in biomedical research especially to culture viruses.
Whilst the patient ultimately died of her cancer eight months after
the operation, her cells have lived on, still surviving in
laboratories today.  HeLa cells have somehow acquired cellular
immortality, in that the normal mechanisms of programmed cell death
after a certain number of divisions have somehow been switched off.
We apply our approach to the HeLa cell  gene expression data of
\cite{fujita}. Data corresponds to $94$ genes and $48$ time points,
with an hour interval separating two successive readings (the HeLa
cell cycle lasts 16 hours). The 94 genes were selected, from the
full data set
 described in \cite{whitfield}, on the basis of the
association with cell cycle regulation and tumor development.

\begin{figure}[ht]
\includegraphics[width=10cm]{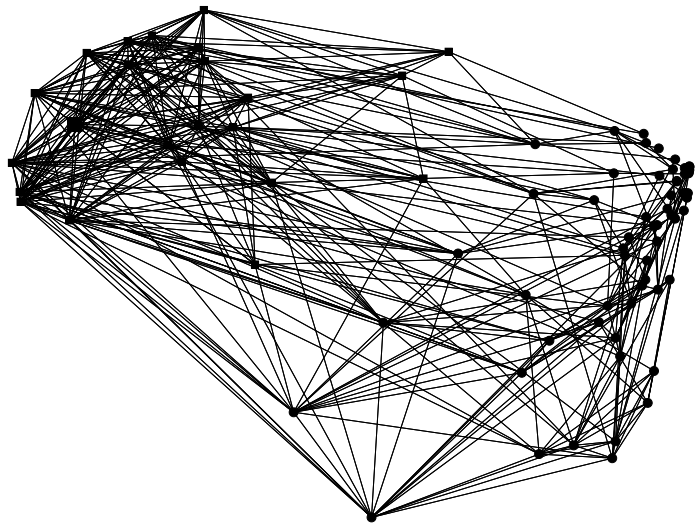}\caption{{\rm The undirected network obtained drawing a link between all pairs of significantly correlated
genes of the HeLa data set (9 genes are not represented here as they
are not correlated with any other gene; hence the number of nodes is
85). Squares and circles corresponds to the partition in two modules
performed by the method described in \cite{boccaletti}. The large
square corresponds to the transcriptional factor NFkB.
\label{fig10}}}\end{figure}

First of all, we perform the analysis of the static pairwise
correlations between  time series: 800 pairs of genes are
significantly correlated. Drawing a link for each correlated pair
leads to an undirected network depicted in figure (\ref{fig10}): it
is clear that there are two modules, and symbols in figure
(\ref{fig10}) corresponds to the partition by the method of module
identification described in \cite{boccaletti}. The first module is
made of 23 genes and corresponds to the regulatory network of the
transcriptional factor NFkB \cite{nfkb}; it contains several well
known activators and targets of NFkB \cite{nfkb2}, like, e.g., A20,
ICAM-1, IL-6,VCAM-1, IkappaBa, JunB, MCP-1, FGF2, Cyclin. The second
module, 62 genes, appears to be orchestrated by transcriptional
factors p53 and STAT3. Note, however, that the two modules are not
 independent, as they form a highly related
network. The proto-oncongene c-myc appears to be central between the
two modules: it has 12 significant static correlations with both
modules. After the discussion in section IV-B, we assume that the
modular structure depicted in figure (\ref{fig10}) is the result of
regulatory mechanisms acting on time scales much smaller than the
sampling time.

Next, in order to detect causalities acting on the time scale of the
sampling time,  we apply bivariate Granger causality analysis. For
all pairs $(i,j)$, we use the linear version ($p=1$, $m=1$) of our
approach and evaluate the Pearson correlation coefficient $r$,
equation (\ref{causalityindex3}), for the causality $i\to j$. Due to
the small number of samples, we do not use t-test to evaluate the
probability $\pi$ corresponding to $r$: we generate a set of
surrogates by permuting the temporal indices of the i-th times
series while keeping fixed those of the j-th time series. The
probability $\pi$ is identified  with the fraction of times that an
higher coefficient is obtained over $3\times 10^6$ random shufflings
of time indices of the i-$th$ time series. Moreover, we use the
False Discovery Rate (FDR) method \cite{fdr} instead of Bonferroni's
correction. FDR works in the following way: the $94\times 93=8742$
Pearson coefficients are ordered, $\{r_\ell\}$, according to their
increasing $\pi_\ell$ values, and a parameter $q$, which controls
the fraction of false positive, is set to $0.05$. The index
$\ell^\prime$ is identified as the largest such that for all
$\ell\le \ell^\prime$ we have $\pi_\ell\le \frac{\ell q}{8742}$.
Pearson coefficients $r_\ell$ are accepted for $\ell\le
\ell^\prime$. This procedure selects 19 causal relationships, out of
8742; they are listed in Table I. IkappaBa is the most abundant
inhibitory protein for NFkB \cite{chen}: our approach detects the
significant causality IkappaBa
 $\to$ NFkB. We find that NFkB is also casually related to IAP, an
anti-apoptotic gene, and B99 (a direct target for transcriptional
activation by p53: here no significant interaction  between B99 and
p53 has been detected). Three causality relationships involve
Bcl-xL, the dominant regulator of apoptosis (active cell suicide)
and TSP1, a peptide shown in some tumor systems to be linked with
angiogenesis. Notably Table I also contains fibroblast growth
factors, FGF7 and FGFR4, the tumor necrosis factor Killer /DR5, the
myeloid tumor suppressor gene PKIG, the tumor protein TPD52-L and
Cyclin E1, a gene which is overexpressed in many tumors. In
\cite{fujita} data have been analyzed with the sparse vector
autoregressive model, a multivariate L1 approach which depends on a
regularization parameter, $\lambda$, fixed by cross-validation. Only
one causality relationship, out of the 19 in Table I, was revealed
also in \cite{fujita}: A20$\to$Bcl-XL.
\begin{center}
\begin{table}
\begin{tabular}{|ccc|}
\hline
IkappaBa & $\to$& NFkB\\
NFkB & $\to$& B99\\
IAP & $\to$& NFkB\\
c-myc & $\to$& FGFR4\\
TSP1& $\to$& c-myc\\
Killer/DR5 & $\to$& c-myc\\
R2 & $\to$& c-myc\\
VCAM-1& $\to$& TPD52L\\
Bcl-XL & $\to$& OCT4\\
A20 & $\to$& Bcl-XL\\
IRF-2 & $\to$& BRCA1\\
TSP1& $\to$& Bcl-XL\\
Cyclin E1 & $\to$& E2F-1\\
OCT4 & $\to$& VCAM-1\\
FGF7 & $\to$& MCP-1\\
TPD52L& $\to$& TNF-a\\
TPD52L & $\to$& MASPIN\\
PKIG& $\to$& ICAM-1\\
PKIG& $\to$& TSP1\\
\hline
\end{tabular}
\caption{Causalities for HeLa gene network.}
\end{table}
\end{center}
\section{Discussion}
Our method of analysis of dynamical networks is based on a recent
measure of Granger causality between time series, rooted on kernel
methods, whose magnitude depends on the amount of flow of
information from one series to another. By definition of Granger
causality, our method allows analysis of networks containing cycles.
Firstly we have demonstrated the effectiveness of the method on a
network of chaotic maps with links obtained assigning a direction to
the edges of the well known Zachary data set, using a nonlinear
kernel: perfect reconstruction of the directed network is achieved
provided that a sufficient number of samples is available.

Secondly we studied a simulated genetic regulatory network. The
results from our method were better than those from DBN approach.
However our performance was strongly dependent on the sampling time,
as it occurred also using DBN method. In this example, use of IP
kernel, with  ($p>2$), or Gaussian kernels did not lead to
improvement in the performance with respect to the linear kernel:
this means that these kernel are not suitable to model the nonlinear
constraint connected to the fact that expressions are confined in
$[0,100]$. Further work will be devoted to the search for kernels
capable to capture this kind of nonlinearity: for a given
application one should choose the proper kernel out of the many
possible classes \cite{shawe}.

Then we considered the case of sparse connectivity and limited data.
Using an example consisting in a linear dynamical network on a graph
grown by preferential attachment, we have shown that there are
instances where the multivariate Granger approach is unfeasible, but
the application of bivariate Granger analysis, to every pair of time
series, leads to better results than those from a method based on L1
minimization. Finally we have analyzed a real data set of temporal
gene expression samples from HeLa cells. The static correlation
analysis between time series, which is the result of regulation
mechanisms with time scaler faster than the sampling rate, revealed
the presence of two modules. Use of bivariate Granger causality has
put in evidence 19 causality relationships acting on the time scale
of one hour, all involving genes playing some role in processes
related to tumor development. Our result on HeLa data has very
little overlap with those from the output of a method based on
multivariate L1 minimization, but this is not surprising, as we
observed the same fact  also on the linear dynamical model of
Section V, where the true connectivity was known. We remark that
currently available data size and data quality make the
reconstruction of gene networks from gene expression data a
challenge.

Detecting cause-effects influences between components of a complex
system is an active multidisciplinary area of research in these
years. The kernel approach  here presented, provides a statistically
robust tool to assess drive-response relationships in many fields of
science.

The authors thank  J. Yu, A. Hartemink  and E. Jarvis (Duke
University, USA) and A. Fuijta, C.E. Ferreira (University of Sao
Paolo, Brazil) for valuable correspondences.

\end{document}